\documentclass[onecolumn,fleqn,usenatbib]{mnras}

\usepackage{newtxtext,newtxmath}
\usepackage[T1]{fontenc}

\DeclareRobustCommand{\VAN}[3]{#2}
\let\VANthebibliography\thebibliography
\def\thebibliography{\DeclareRobustCommand{\VAN}[3]{##3}\VANthebibliography}

\newcommand{\HI}{\hbox{\rmfamily H\,{\textsc i}}}
\newcommand{\HIsub}{\hbox{{\scriptsize H}\,{\tiny I}}}
\newcommand{\MHI}{\hbox{$M_{\HIsub}$}}
\newcommand{\msun}{\hbox{${\rm M}_{\odot}$}}
\newcommand{\kms}{\hbox{km\,s$^{-1}$}}
\newcommand{\askapsoft}{ASKAP{\sc soft}}

\newcommand{\New}[1]{{\bf\color{red}{#1}}}
\renewcommand{\New}{}
\newcommand{\Old}[1]{{\color{pink}{#1}}}
\renewcommand{\Old}{}

\usepackage{graphicx}	% Including figure files
\usepackage{amsmath}	% Advanced maths commands
\usepackage{graphicx}
\usepackage{lscape}
\usepackage{color}
\usepackage{amsmath}
\usepackage{multirow}
\usepackage{booktabs}
\usepackage{caption}
\usepackage{subcaption}
\usepackage{ulem} % or \usepackage[normalem]{ulem}
\usepackage{parskip}
\usepackage{makecell}

\graphicspath{{./}{Figures/}}

\title[Compression Options for radio interferometry]{Options for Compression of radio interferometry data: lossy
  compression of visibilities and lossless compression of uv-visibility grids for the MHONGOOSE survey
}

\author[R. Dodson et al.]{
Richard Dodson$^{1}$\thanks{E-mail: richard.dodson@icrar.org}, Jonghwan Rhee$^{2,1}$, W.J.G. de Blok$^{3,4,5}$, Martin Meyer$^{1}$, 
\newauthor Alexander Williamson$^{1,6}$,  Daniel Mitchell$^{7}$,  Krist\'of Rozgonyi$^{1}$, Mar\'ia J. Rioja$^{1,8}$, Pascal J. Elahi$^{9}$\\ 
$^{1}$International Centre for Radio Astronomy Research (ICRAR), University of Western Australia, 35 Stirling Hwy, Crawley, WA 6009, Australia \\
$^{2}$Australia Telescope National Facility, CSIRO Space \& Astronomy, P.O. Box 1130, Bentley, WA 6102, Australia \\
$^{3}$Netherlands Institute for Radio Astronomy (ASTRON), Oude Hoogeveensedijk 4, 7991 PD Dwingeloo, The Netherlands\\
$^{4}$Dept. of Astronomy, Univ. of Cape Town, Private Bag X3, Rondebosch 7701, South Africa\\
$^{5}$Kapteyn Astronomical Institute, University of Groningen, PO Box 800, 9700 AV Groningen, The Netherlands\\
$^{6}$Australian SKA Regional Centre (AusSRC), University of Western Australia, 35 Stirling Highway, Crawley, WA 6009, Australia\\ 
$^{7}$Australia Telescope National Facility, CSIRO Space \& Astronomy, P.O. Box 76, Epping, NSW 1710, Australia\\
$^{8}$Observatorio Astron\'omico Nacional (IGN), Alfonso XII, 3 y 5, 28014 Madrid, Spain\\
$^{9}$Pawsey Supercomputing Research Centre, CSIRO Space \& Astronomy, P.O. Box 1130, Bentley, WA 6102, Australia \\
}

\date{Accepted XXX. Received YYY; in original form ZZZ}

\pubyear{\the\year{}}

\begin{document}
\label{firstpage}
\pagerange{\pageref{firstpage}--\pageref{lastpage}}
\maketitle

\begin{abstract}
\Old{Next generation radio astronomy telescopes are challenging existing data reduction paradigms. With ever more antennas, larger bandwidths, and  sometimes multiple primary beams, they often generate more observed data products than can readily be stored long-term.  Thus, data storage becomes a major cost driver and processing constraint.  In this paper, we test two methods of addressing this problem: grid-stacking, a two-stage lossless compression solution; and the lossy compression of the raw visibilities before traditional processing. To demonstrate these solutions we utilised a deep imaging pipeline based on software for the ASKAP telescope, \askapsoft{}, but applied to a strong source (NGC\,1566) from the deep MeerKAT \HI{} spectral line project, MHONGOOSE.  The grid-stacking solution reproduces the spectrum from traditional processing to within better than 0.7\%, and also allows for the reconstruction of other weighting scales without significant computing costs. In comparison, image-stacking also reproduces the spectrum from the traditional processing, to within better than 3\% but with worse image residuals in the cube. The lossy compression, even at a near ten-fold reduction in file size, reproduces the spectra almost perfectly (to better than $\sim$0.01\% in all cases). Thus both compression methods are promising solutions, and we discuss considerations for their application.}

\end{abstract}

%\keywords{Radio interferometry, Data compression, Uv-gridded data, Lossy compression, Spectral line surveys}

%%%%%%%%%%%%%%%%%%%%%%%%%%%%%%%%%%%%%%%%%%%%%%%%%%

%%%%%%%%%%%%%%%%% BODY OF PAPER %%%%%%%%%%%%%%%%%%

\section{Introduction} \label{sec:intro}

% future of radio astronomy/telescopes
We are entering an exciting era in radio astronomy, characterised by an unprecedented expansion in observational capability, driven by at least three forthcoming facilities: the Square Kilometre Array \citep[SKA][]{Dewdney:2009}, the next-generation Very Large Array \citep[ngVLA][]{ngvla_reference_design}, and the Deep Synoptic Array \citep[DSA][]{dsa}.
Collectively, these instruments will deliver approximately an order-of-magnitude increase in instantaneous bandwidth and several orders-of-magnitude enhancement in effective collecting area, and hence sensitivity, relative to current state-of-the-art radio telescopes. This substantial improvement will enable high-fidelity large-area surveys of the radio sky, the detection of the redshifted 21-cm signal from the epoch of reionization corresponding to the formation of the first stars, and precise measurements of spectral line and continuum emission from millions of galaxies.

Realising the scientific potential of these instruments requires the radio astronomy community to handle and process data volumes that are without precedent in the field. The anticipated expansion in data throughput is the dominant cost driver for the high-performance computing infrastructures needed to support next-generation observatories. For the SKA, the data rate emerging from the correlators is expected to be on the order of 1\,TB\,s$^{-1}$, which must be buffered locally and processed in near real time due to limited long-term storage capacity \citep[see, e.g.][]{ska-overview}. This constraint will have the most pronounced impact on deep, multi-epoch observations, where conventional imaging workflows involve iterating over the full raw visibility data set for each major deconvolution cycle.  Meeting this challenge will require the development of new methods to reduce storage and processing costs, while still delivering science images of sufficient quality. %(need to word this better)
{For Australian Square Kilometre Array Pathfinder \citep[ASKAP,][]{Hotan:2021}, and SKA, traditional imaging methods require more storage than can be managed with the cost-constraints, thus both instruments will require improved processing solutions to complete their deep \HI{} surveys.} 

Radio interferometric data presents specific challenges. A large fraction of the measured visibilities are noise-dominated because the signal from any spatially localised region of high surface brightness is distributed across all baseline–frequency–time samples. Moreover, the radio sky is intrinsically sparse, and the brightness temperature of astrophysical sources, T$_{\rm sky}$, is generally much lower than the system noise dominated by the receiver temperature, T$_{\rm rec}$. These characteristics open up several avenues for more efficient data handling and processing methodologies, and
%which we will investigate in the following sections.
% summary of progress to date
%The 
the main thrust of our recent investigations is how to address this challenge, for which we are using the SKA pathfinder instruments and observational projects.

Our recent work has been exploring the requirements for alternative data products, with particular emphasis on the comparison of traditional and image-stacked data products with \textit{uv}-grids and direct visibility data compression. 
The goal is to reduce both the \Old{peak} computation and storage demands while preserving scientific fidelity.
In \citet{dodson_25} we investigated lossy compression strategies using simulated and real radio interferometric datasets to assess their impact on imaging quality and scientific integrity. 
This work demonstrated that carefully designed compression schemes can substantially reduce data volumes while maintaining acceptable image fidelity, achieving compression factors of as much as an order of magnitude with negligible degradation to key scientific metrics. 
These findings suggested that compression may provide a viable pathway to reducing storage and processing requirements for future deep imaging experiments.
In \citet{williamson_25} we investigated the computational considerations for Deep Investigation of Neutral Gas Origins \citep[DINGO,][]{Meyer:2009,Rhee:2023} and ASKAP of storing the data averaged to the \textit{uv}-grids, measuring the I/O performance, the introduced errors as a function of error bound and the impact on the imaging. 
In \citet{rhee_25} we demonstrated that, for DINGO and ASKAP, summing daily (8h) \textit{uv}-grids to produce the final image cube gave better results than summing the daily restored images, and the former gave a result that was closer to that of traditional processing.
These two papers explored the alternative deep imaging workflows using approximately 200 hours of ASKAP observations. 
Our conclusion were that \textit{uv}-grid approaches can reproduce traditional deep imaging results while substantially reducing data volume requirements and spreading the compute requirement over the time span of the observations. 

Building upon these earlier investigations, the present work aims to extend and unify these approaches in several key ways. 
Firstly, we evaluated these alternative imaging methodologies using an independent deep imaging dataset from the Meer-Karoo Array Telescope \citep[MeerKAT,][]{meerkat}, the other SKA pathfinder, thereby testing the generality and robustness of the earlier conclusions and also the \askapsoft{} codebase. 
Secondly, we performed a direct comparison between direct compression-based approaches and \textit{uv}-grid methods, 
allowing a quantitative assessment of their relative performance, limitations, and scientific fidelity within a common analysis framework. 
Finally, we investigate the feasibility of reweighting strategies within these reduced data representations, exploring whether the flexibility that is typically available when imaging raw visibilities, can be retained in the gridded format.

\section{Data}

\subsection{The MeerKAT {Precursor}}
The MeerKAT radio telescope \citep{meerkat} is the main SKA-Mid \citep{Swart:2022} precursor, located in the radio-quiet Karoo region of South Africa. 
The array comprises 64 offset-Gregorian parabolic dishes, each 13.5m in diameter, arranged with a compact, dense core and a distribution of longer baselines extending to roughly 8\,km. 
This configuration delivers excellent surface-brightness sensitivity on scales of tens of arcminutes to a few arcseconds, making MeerKAT
well suited to both deep, low-column-density imaging and high-resolution studies of individual galaxies.
MeerKAT receivers cover roughly 0.58–3.5\,GHz (the UHF through to S-band windows), using sensitive single-pixel feeds and a modern digital backend and correlator that support wide instantaneous bandwidths, flexible spectral setups, and high dynamic range imaging. 
The combination of large collecting area, low local radio-frequency interference, and an optimised array layout yields exceptional sensitivity to faint \HI{} emission and diffuse continuum structure.

Thus, one of the driving science goals for MeerKAT is deep extragalactic \HI{} and continuum surveys to trace galaxy evolution, gas accretion, and the low-column-density cosmic web that embeds galaxies. 
In recognition of this, a number of programs were selected to be MeerKAT Large Legacy Science Projects, to exploit the array’s sensitivity and imaging fidelity for deep \HI{} work. These include surveys equivalent to DINGO: LADUMA \citep[Looking At the Distant Universe with the MeerKAT Array,][]{Holwerda:2012, Blyth:2016}, MFS \citep[MeerKAT Fornax Survey,][]{mfs} and also the pointed survey MHONGOOSE \citep[MeerKAT \HI{} Observations of Nearby Galactic Objects: Observing Southern Emitters,][]{mhongoose}.
These programs combine long integrations, careful calibration strategies, and advanced imaging pipelines to push the observable \HI{} column-density limits and to characterise gas kinematics and environment across a wide range of mass and redshift.
For the present study we accessed the fully processed data products delivered by the MHONGOOSE team, taking advantage of their multiple epochs to test the performance of different analysis methods on a system that most closely mimics the SKA-Mid.

\subsection{The MHONGOOSE Survey} \label{subsec:mhongoose}

The MHONGOOSE survey is a deep \HI{} survey pathfinder conducted with MeerKAT \citep{mhongoose}, designed to investigate gas accretion and star formation processes across a broad range of galaxy properties. 
The survey was constructed to sample a representative range of \HI{} masses, stellar masses, star formation rates, and rotation velocities (and, by implication, dark matter halo masses), using a sample of 30 galaxies located within a distance of {23} Mpc. 
This sample enables the study of the spatial distribution and kinematics of neutral hydrogen over approximately five orders of magnitude in \HI{} mass. 
Owing to its highly sensitive pointed observing modes, MeerKAT is particularly well suited to the detection and detailed imaging of faint gaseous components accreting onto galaxies from their circumgalactic and intergalactic environments. 
An additional objective of the survey is to characterise the low-\HI{}-column-density structures in the vicinity of the target galaxies.
For this reason the project imaged the data at different resolutions that have optimal sensitivity to different scales. 

The MeerKAT MHONGOOSE observations {took place from October 2020 to December 2024 with observing times of 55h per galaxy. The highest achievable resolution is $\sim7^{\prime \prime}$ , 
with channel widths of 6.5kHz (after two-channel binning from the original 3.3 kHz spectral resolution) and correlator integrations of 8.1s.} 
The observations of all the galaxies were broken into 10 approximately equal epochs, which means the final image is significantly deeper than those of any single day (i.e., any of the individual 5.5h observations).
Images were made at six different resolutions, by varying the weighting scheme \citep[see Table 4 in][]{mhongoose}.
Thus this is an extremely interesting dataset for the investigation of the next generation of data processing methods, as we will be able to study the difference between processing in a single run and over several individual runs, explore solutions for compressing the raw data and changing the weighting schemes on the fly.

\section{Methods} \label{sec:method}

All of our imaging was undertaken with minor adjustments to the basic \askapsoft{} spectral line processing \citep{Cornwell:2011, Whiting:2017, Wieringa:2020, whiting2020asp}.
Traditional processing, where one retains the raw data in the visibility domain, will not be possible with the storage limitations for the next-generation instruments, thus evaluation of the impact of the alternative strategies that we will be forced to use is imperative. 
Nevertheless, with the current data volumes, we can use traditional methods to provide a reference result and explore other approaches.
The alternative strategies are: 
the default option is  to image the daily data independently and to store and then stack these daily images, which leaves the final residuals containing a significant amount of uncleaned flux; 
the grid-stacking option is to store and then stack the daily deconvolved visibility grids, which allows for a final deep minor CLEAN cycle but no major cycles; 
the lossy compression option is to store the datasets in a compressed format and then image those in the traditional fashion, which has impact on the image quality that is only now being explored. 
The image-stacking is now widely accepted to be a poor option; even with the pathfinders we can see that it leads to difficulties in the analysis due to the mixture of cleaned and uncleaned flux in the measurements \citep[for example discussions in][]{wallaby_performance} and image imperfections \citep[e.g. Figure 3 in][]{rhee_25}. 
The grid-stacking approach was not immediately supported by the software, nor was it carefully tested for the subtle systematic impacts that would undermine the purpose; this has now been addressed \citep{Rozgonyi:2021}.
Regarding the option of lossy compression, the \HI{} community remains cautious: any adverse effects would only become apparent after thousands of hours of observation and once introduced would be irreversible. A study such as \citet{Rozgonyi:2021} has not yet been performed to test the impacts. 

To address the first of these approaches for deep \HI\ surveys we have published an analysis using the other SKA pathfinder project on a SKA telescope site, DINGO with ASKAP \citep{Rhee:2023}, with which we were able to make a detailed comparison of the first three (traditional processing, grid-stacking and image-stacking) of these methods \citep{rhee_25}. 
%Traditional processing will not be possible in the SKA-era, and image-stacking introduces image residuals.
The grid-stacking approach is very compatible with compression \citep[as many \textit{uv}-cells hold no data and have value zero; ][]{williamson_25} and is an example of optimal baseline dependent averaging (as that is implicit in the gridding process). It allows for a final deep deconvolution, albeit limited to a single Hogbom CLEAN as major cycles are not possible, and this is what we are favouring for the DINGO survey. % currently.
%
%We have already studied the impact of grid-stacking compared to traditional processing for the ASKAP deep HI survey, DINGO \citep{rhee_25}.
To demonstrate that this approach is valid for MeerKAT, and this offers a viable path for SKA, we extended these studies to the MHONGOOSE survey. 
We also included the alternative approach of working directly from compressed datasets using traditional methods \citep{dodson_25}. 
In this case, we  used ADIOS and the various lossy compressors that it offers (specifically SZ\footnote{https://szcompressor.org/} \citep{sz} and MGARD \citep[Multi-Grid Adaptive Reduction for floating-point scientific Data,][]{Gong:2023}), where the user sets an adjustable data-based error-bound as the limit to compression which they determine is acceptable. 
We compared this to the approach of DYSCO \citep[DYnamical Statistical COmpression,][]{dysco}, which sets an adjustable limit on the number of bits used. 
%The data provided to this investigation was flagged, self calibrated and continuum subtracted, following the description in \citet{mhongoose}. 

The dataset used for this investigation was a single galaxy from the MHONGOOSE survey, 
NGC\,1566 (J0419--54), and we applied the \askapsoft{} processing to 200 channels of the already calibrated, 
flagged, continuum subtracted \HI{} dataset, following the description in \citet{mhongoose}. 
{The 200 channels spanned 1412.6 to 1413.9\,MHz, or 1380 to 1650\,\kms.}
NGC\,1566  is {among the nearest bright} Seyfert galaxies, and is also known as the Spanish Dancer. 
The proximity of the galaxy, along with stunning and clearly marked face-on spiral arms and a supermassive black hole at the core, have made it the subject of much scientific study in the astronomy community. {An example study of the MHONGOOSE data of NGC 1566 is presented in \citet{Maccagni:2024}.}
% solutions
%  Lossy: dysco 
%  Lossy: adios 
%  Lossless: grid-stacking

%

\subsection{The lossless grid-stacking solution}

The grid-stacking solution was first introduced as the \texttt{msuvbin} tool in CASA \citep{msuvbin}. 
It has been further developed as part of the PhD thesis of \citet{Rozgonyi:2021} and with the shared ICRAR-Pawsey-AusSRC HiVIS\footnote{\HI\ Visibility Imaging for the SKA} PaCER\footnote{Pawsey Centre for Exascale Readiness} project \citep{Rozgonyi:2022}. 
It has the advantage of enabling minor-cycle CLEANing at full survey depth and the improving model subtraction.  
However, since the daily data are stored after application of the \textit{uv}-kernels and combined on the 2D \textit{uv}-plane, major cycles are not possible, so we must ensure that we work with nearly final residuals, as in that case further major cycles are not required.
The generation of daily residual grids can be part of the standard ASKAP processing using \texttt{imager} with limited additional costs, where the cubes are CLEANed with 3 major cycles.
However, some additional steps are required. 
Firstly, the daily \textit{uv}-grid residuals must be stacked, imaged and deconvolved, for which we use the \askapsoft{} deconvolution command, \texttt{cdeconvolver}.
Following this, the daily CLEAN models must be averaged (taking into account the per channel weights per epoch), convolved with the `clean' or restoring beam and added back into the final data once the final residual cube is generated in the same manner as the final model components, albeit in a separate processing step.
We found that the averaging of the beam in traditional imaging, or by either of the stacking approaches, led to slightly different restoring beams. 
These are of the order of a percent, and would not be noticeable, other than for the fact we are making a direct comparison between different methods and a relative `truth' can be defined. 
Fortunately, we can define a shaped beam in \askapsoft{} where the tapering is adjusted per channel and per epoch, to take into account the data sampling to produce an identical resolution for all epochs and channels. 
DINGO uses this function, for example, to provide a common resolution across wide-bandwidths.
The restored beam for this project was that of the shaped beam in the deconvolution; 20\arcsec{} for Robust 0.5 and 40\arcsec{} for Robust 2.0.
Neither of these steps took significant compute-time. 
We combined the per epoch restored images to provide a reference point as well. 
These were averaged with the per epoch per channel weights, as were the residual images.
In this case, the restoring beam was the weighted averaged restoring beam across the epochs. If the the restoring beam is shaped this is the same as in the inputs, but is important if beam-shaping is not performed. 

% The PCF grid represents the baseline sampling of the system, as the Wiener filtering requires the count of the samples per cell without kernel weighting. 

The data retained for the grid-stacking also allows for the alteration of the weighting scheme.
For the DINGO project \askapsoft{} \texttt{imager} in addition to the daily image produces: a residual visibility grid, a PSF (Point-Spread Function) grid, a PCF (Pre-Conditioning Function) grid and the spectral model components subtracted to produce the residual visibilities, which can be combined in \texttt{cdeconvolver}. The visibility grid is a grid representation of the 3-dimensional visibilities projected onto a 2-dimensional grid for each frequency channel (producing a 3-dimensional grid). 
The PSF grid represents the size, location and weighting of the complex-valued convolutional kernels applied during the gridding process. 
The  PCF grid represents the baseline sampling of the system, as the Wiener filtering requires the count of the samples per cell without kernel weighting. 
Thus these are sufficient to change the weighting and tapering, which is an option that can be included at the \texttt{cdeconvolver} stage.
% These grids are passed to the \texttt{cdeconvolver} application which, if more than one observation is provided, will sum the grids together as they are read in. These summed grids are then imaged for analysis. Here we validate that the resulting image is free of detrimental RFI and that the sensitivity of the image is not degraded.

\subsection{The lossy compression solution}

It has long been recognised in radio astronomy that the Gaussian white-noise distribution of nearly all visibility does not allow for significant compression, without allowing for losses in precision.
The traditional approach to reduce the volume was to average the data, but normally the correlator setup maximises the averaging that can be achieved without loss across the field of view. 
A refinement is to implement a baseline dependent averaging (BDA), by recognising that the channel width (for continuum targets) and the time sampling implemented at the correlators are those for the longest baselines. 
Consequently, one could have used greater averaging for the shorter baselines in the dataset. 
BDA is a post-correlation solution that introduces further averaging for these shorter baselines, in time and frequency (if not a spectral line dataset). This has been shown to have good results \citep{bda_stefan}, but can cause difficulties for down-stream processes, as the data sampling is no longer regular. 
An alternative is to lossy-compress the data, without impacting the average value, in an intelligent fashion to minimise the impact. %, i.e. baseline dependent compression (BDC).
We have focused on multi-grid compression techniques, that guarantee that the average is not affected as they generate residuals around a running mean. %\citep{Gong:2023} 

The first implementation of lossy compression in Radio Astronomy (to our knowledge) was DYSCO \citep{dysco},  which can reduce the number of bits in which the values are stored whilst limiting the impact on the data quality and was developed for the field.
\Old{DYSCO first normalises the data before quantisation with dithering, with a range of normalisation options. We have used the default method of a 2.5$\sigma$-truncated Gaussian distribution and AF-normalization.}
  % Adminfo={ "TYPE": "DyscoStMan", "NAME": "dysco", "SPEC": {
  %                         'dataBitCount': int(bitcount),
  %                         'weightBitCount': 12,
  %                         'distribution': 'TruncatedGaussian',
  %                         'normalization': 'AF',
  %                         'studentTNu': 0.0,
  %                         'distributionTruncation': 2.5
  %                     }
The second demonstration (to our knowledge) was to use multi-grid methods such as SZ or MGARD \citep[e.g.][]{dodson_25}, made available via ADIOS, 
where one specifies the allowable error rather than a fixed bit level. 
\Old{From the compressors supported in ADIOS we prefer SZ \citep{sz}  and  MGARD \citep{Gong:2023}, as they do not use block-based mathematical transforms, which introduces correlations between data points that can be detected in the resultant images. 
Both compressors were developed for generalised scientific data and offer error-controlled lossy compression rooted in multi-grid theories. They transform floating-point scientific data into a multilevel representation, followed by quantisation, lossless encoding, and ultimately generating a self-describing compressed buffer. 
MGARD uses a multilevel, multigrid and finite-element model during compression over specific norms and quantities of interest and is mathematically guaranteed  by finite element theories.
SZ uses a level-wise hybrid interpolation predictor to provide better compression ratios and lower distortions.
The former is specifically designed for GPUs, although there is a CPU implementation. The latter has a significantly faster CPU implementation, and also a GPU version. We have limited our investigations to the CPU implementations, as the compression time was not significant compared to the imaging.
}

In noise-limited data these alternatives have virtually the same performance, but we prefer to be able to specify the limit in the data units, as for in radio astronomy we know the precision of the measurement from the SEFD. 
If the noise introduced by lossy compression is insignificant compared to the SEFD the sensitivity is not impacted; this is harder to guarantee for DYSCO.
Furthermore, multigrid methods encode the trend as well as the mean value, which gives a more accurate reconstruction.

For the lossy compression leg of the study the data column was compressed with DYSCO, to a fixed fraction of the size (a quarter or an eighth), or with SZ/MGARD, to an error bound of a fraction of the SEFD ($\sim$2\% or $\sim$20\%). These both produced a similar level of compression and are beyond what we would normally recommend for imperceptible impacts, but are ideal to demonstrate the limit of the achievable acceptable performance.
By working with compressed data we have not altered the data sampling, so all subsequent data handling is unchanged, as long as the core libraries support the compression method (i.e. DYSCO or SZ/MGARD under ADIOS). Both of the latter are standard CASACORE\footnote{https://github.com/casacore/casacore} options, so this is not unduly challenging. See github.com/ICRAR/docker-casacore/ for an example of python code to handle these columns.
We note that, to be able to use the standard tools on setonix without recompiling, we uncompressed the data to a tiled storage managed column in the measurement set, which captures the impact of the compression but doesn't require a bespoke compilation. This also avoided some bottlenecks we had with pseudo-random access of a compressed ADIOS datacolumn
which was ordered by time, which is an issue for imaging tools. 
This is not an issue for the serial access pattern we have for reading and writing the \textit{uv}-grids in \askapsoft{}.

\subsection{Source Finding with SoFiA}

The next step is to conduct {\HI} source finding with the data products from the three methods and compare the derived physical properties of NGC\,1566, the target {\HI} source in the field and J0422--5455, a much weaker source that also falls within the frequency coverage. For this analysis, we use an automated 3D source finding software dedicated to {\HI} surveys, called Source Finding Application \citep[SoFiA,][]{Serra:2015a, Westmeier:2021}. It produces masks, moment maps and {\HI} spectra of detected sources with their physical parameters derived, such as positions (or coordinates), {\HI} velocity (or redshift), fluxes (peak and integrated), and velocity widths (e.g., $W_{20}$ and $W_{50}$). The same mask, generated in the traditional analysis, was used for all the spectral analysis. %\EB{(Maybe some additional info here on kernels used etc?)}
{We used the default smooth-and-clip (S+C) finder algorithm in SoFiA, which smooths the data across multiple spatial and spectral scales. The specific configuration parameters used were scfind.kernelsXY = 0, 5, 10, scfind.kernelsZ = 0, 3, 7, 15, 31, and a detection threshold of scfind.threshold = 4.}% A complete overview of the SoFiA parameters is provided in the Appendix.}

\section{Results}

\subsection{Comparison of imaging with WSClean and {\askapsoft}}

We re-imaged the data with the same parameters as in \citet{mhongoose} using WSClean, and with similar parameters as in \citet{rhee_25} using \askapsoft{} imager.
There is a definitional difference between \askapsoft{} and WSClean, with Stokes $I=XX+YY$ for the former and $I=(XX+YY)/2$ for the latter; see Figure \ref{fig:ratio_flux} for the demonstration.
For data both calibrated and imaged within the same system the outcomes would be equivalent. However, with data that were calibrated within one system and imaged within the other we had to manually correct for this factor. 
After this correction, the image noise as measured by the standard deviation was $\sim$1\% higher in the \askapsoft{} image, whereas the pixel fluxes was $\sim$10\% higher. We believe that this improvement in the SNR is due to the different imaging modes in \askapsoft{} and WSClean. The former is a multi-scale CLEAN (with scales of 0,6,15 and 30 pixels) whereas the WSClean was without multi-scale CLEAN.
Further experiments where we enabled multi-scale CLEAN on WSClean reduced the differences, but did not remove them. 
We leave a detailed comparison of the two packages for future investigations, 
but note that the compute time for traditional imaging with WSClean was significantly less than that for \askapsoft{} imager.

\subsection{Comparison of grid-stacking to traditional imaging and image-stacking}
%% Referee: Section 4.2.: Section 4.2 could probably benefit from some compression, it is very wordy.

Grid-stacking of the daily datasets, followed by a final minor cycle, is the solution for deep imaging that we have adopted for the DINGO project.
For this MeerKAT investigation we used the \askapsoft{} codebase and followed the same procedure used for the DINGO analysis in \citet{rhee_25}.
The analysis was run on a single node on the Pawsey cluster, setonix, with 20 CPUs per thread and 10 worker MPI instances and 1 master. In all cases we generated the restored image, model, residuals and weights, along with the visibility, PSF and PCF grids. 
We used 3 major cycles with 10,000 iterations, with a MultiScale CLEAN (on 0, 6,  15 and 30 pixels), WProjection and 2048$\times$2048 pixels of 2\,arcsec.
The daily imaging took $\sim$2-3 hours, and the joint deconvolution of all the data took $\sim$30 hours. 
The final stages of a minor CLEAN of the stacked grids and the cube reassembly took a few minutes with 200 threads (one per channel).
The restored beam was set to match those of the traditional processing. 

{\askapsoft{} allows one to shape the beams to a consistent ellipse during imaging, that is to apply an individual taper for each epoch and frequency channel such that a specific restoring beam is returned. 
Initially, we attempted to image the data without shaping the beams, to be consistent with the MHONGOOSE analysis. 
However, the difficulty of precisely specifying the restoring beam when comparing the spectra and the moment-0 images made under the different schemes led to small fractional differences between the analysis methods, even though we expected identical outcomes. 
If we were not carefully comparing the results between methods these data products would have been acceptable, however, because we needed perfectly matching resolutions to detect any meaningful differences, the beam shaping became essential. 
This was mainly an issue for the finest resolution with Robustness 0.5 (i.e. where a few arcsecond error in the specification of the beam would have a greater impact on the reconstruction).
Thus, for the comparison between traditional processing and the two stacking methods, we imaged the data at a fixed tapered resolution for both Robust 0.5 and 2.0. 
Table \ref{tab:res_summary} lists the image beam parameters for all products}.

\begin{figure}
    \centering
    \includegraphics[width=0.8\linewidth]{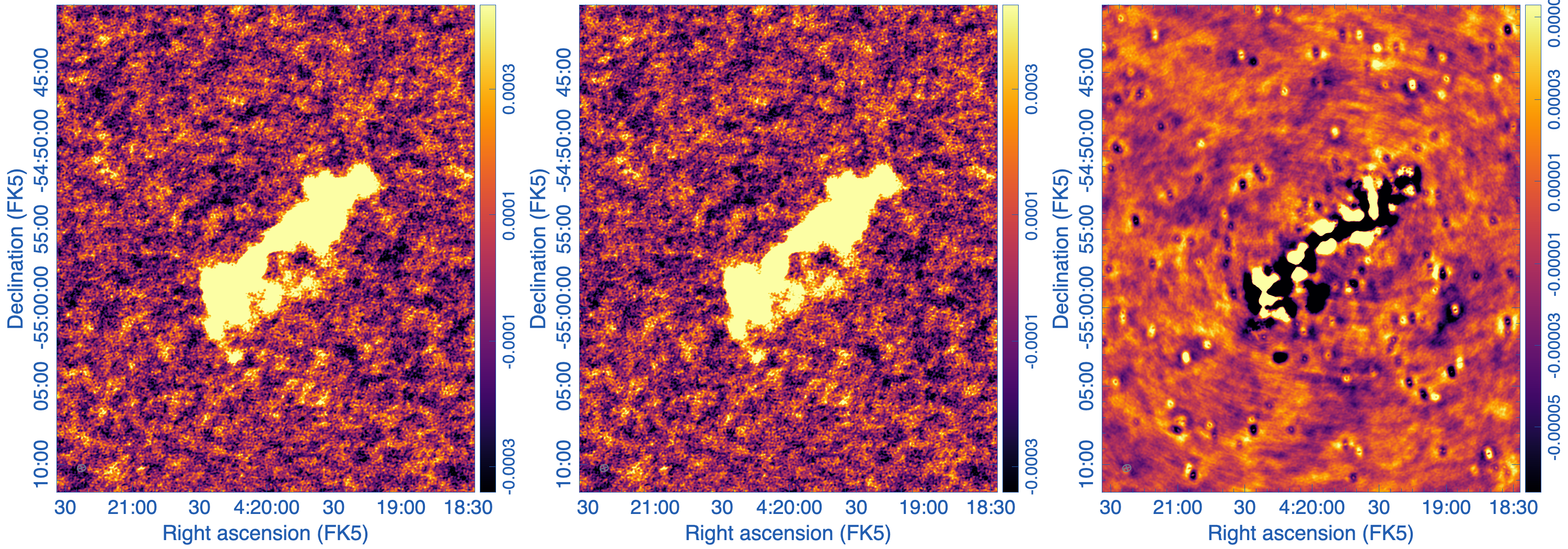}
    %restored.J0419_r20_t40.J0419_r20_t40.Cde-restored.diff_grid_r20_t40.png}
    \caption{{Comparison of the naturally weighted velocity channel at 1509\,\kms{} between Traditional processing (left) and Grid-stacking (middle) with the difference (right). 
    The flux range for the images is 0.4 to 10 mJy/beam and for the differences is $\pm$0.2mJy/beam.}}
    \label{fig:channelmap}
\end{figure}

\begin{table}
    \centering
    \begin{tabular}{r||r|l|l||l}
        Robustness & Traditional/Grid-stacking & Image-stacking & Tapered Resolution & Residual RMS\\% Trad/Grid
         \hline
        2.0 & 32.0$\times$25.4\arcsec{} at -71.5\degr{} & 34.5$\times$24.8\arcsec{} at -42.8\degr{} & 40.0$\times$40.0\arcsec{} at 0.0\degr{}&0.150\\%0.144\\
        1.5 & 31.1 $\times$ 25.4\arcsec{} at -71.0\degr{} & -- & -- &0.171\\%0.144\\
        0.5 & 12.9$\times$10.7\arcsec{} at -58.1\degr{} & 15.7$\times$10.0\arcsec{} at -38.8\degr{} & 20.0$\times$20.0\arcsec{} at 0.0\degr{}&0.175\\%0.166\\
       -2.0 & 6.4 $\times$ 6.0\arcsec{} at -83.0\degr{} & -- & -- &0.255\\%0.257\\
    \end{tabular}
    \caption{Restoring Beams for the image cubes made under various parameters, along with the residual RMS in the traditional processing.}
    \label{tab:res_summary}
\end{table}

\begin{figure}
    \centering
    \includegraphics[width=0.9\linewidth]{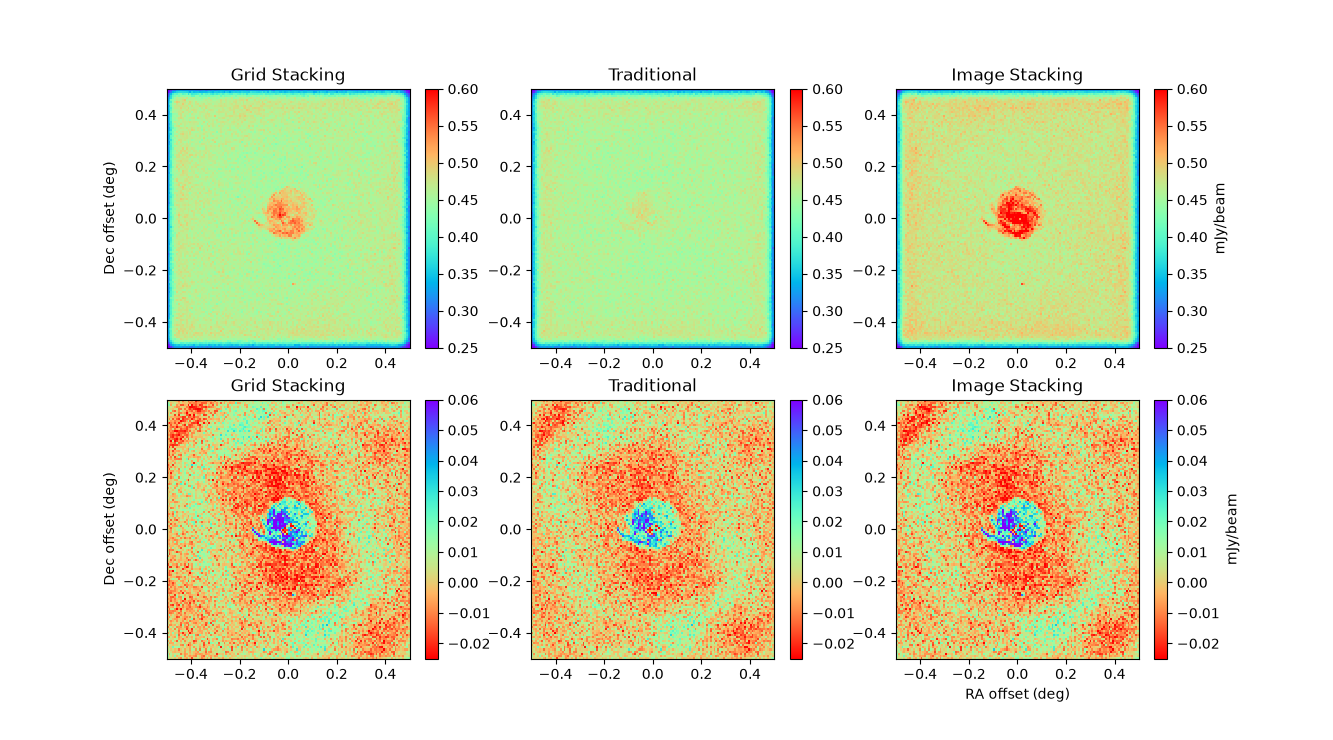}
    \caption{The comparison of the maximum and mean of the residuals left in all of the imaging approaches, with Robust 0.5 and without beam shaping.
    There is no significant flux left in any of the channels for the traditional approach, where as for both grid- and image-stacking there is uncleaned flux remaining from the target. 
    The image-stacking residuals are greater than the grid-stacking residuals, as no final minor CLEAN was applied. 
    The mean of the residuals (to detect weak imaging artefacts) finds similar results, with the artificial ring and the negative bowl around the target being greater in both grid- and image-stacking. %{\bf swap x axis}
    \label{fig:max_res_image}}
\end{figure}

{Figure \ref{fig:channelmap} shows a single 6.5kHz-width channel at 1509\,\kms{} for traditional and grid-stacking for the Robust 2.0 processing, which shows a maximum per pixel difference of $\sim$2\%. The image-stacking channel comparison is very similar.
The majority of the differences are in the detection of non-physical components outside the source, which implies we could improve our results by limiting the CLEAN with a mask, but this would have no consequence for our investigations.
The impact of the differences, being positive and negative components around the edges of structure, are to marginally alter the width of any components whilst leaving the total flux unchanged, and thus not alter the science outputs.
}
Figure \ref{fig:max_res_image} shows the maximum and the mean of the residuals over the cube for the three approaches. 
The maximum per channel residual in the traditional images was at a few times the noise-level, as would be expected. The grid-stacking compared to the traditional processing has more residual flux remaining as no major cleaning cycle could be performed. 
The image-stacking has significantly more, as no minor nor major cleaning cycle could be performed.
The mean of the residual picks up the uncorrected and false structures; that is the ring of positive emission and the negative bowl around the target. In this case the grid- and image-stacking are approximately equal, and both are worse than those for the traditional processing. 

\begin{figure}
    \centering
    \includegraphics[width=0.8\linewidth]{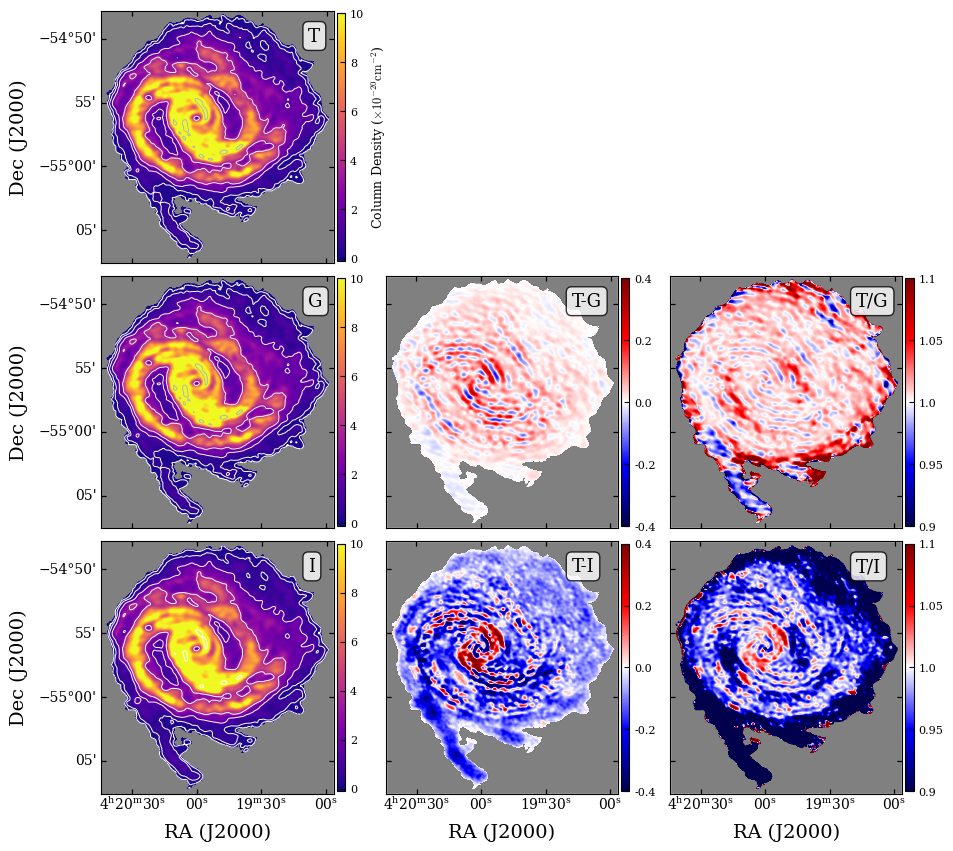}
    \caption{Moment 0 images at 20\arcsec{} resolution and Robust 0.5,  of NGC\,1566 from SOFiA, using (left column) traditional processing (T), Grid-stacking (G) and Image-stacking (I). The middle column is the difference between traditional and the other imaging methods. The right column is the ratio. 
    The moment-0 maps all are visually similar, {but there are some differences in the high spatial frequencies, as indicated by the differences and ratios. Image-Stacking shows a more significant shortfall in recovered flux compared to Grid-Stacking.}
    %The brighter emission on the arms is slightly over cleaned and the intra-arm area is under cleaned.
    %\New{Now Redone with shaped beams}
    \label{fig:image_compare_0.5}}
\end{figure}

\begin{figure}
    \centering
    \includegraphics[width=0.8\linewidth]{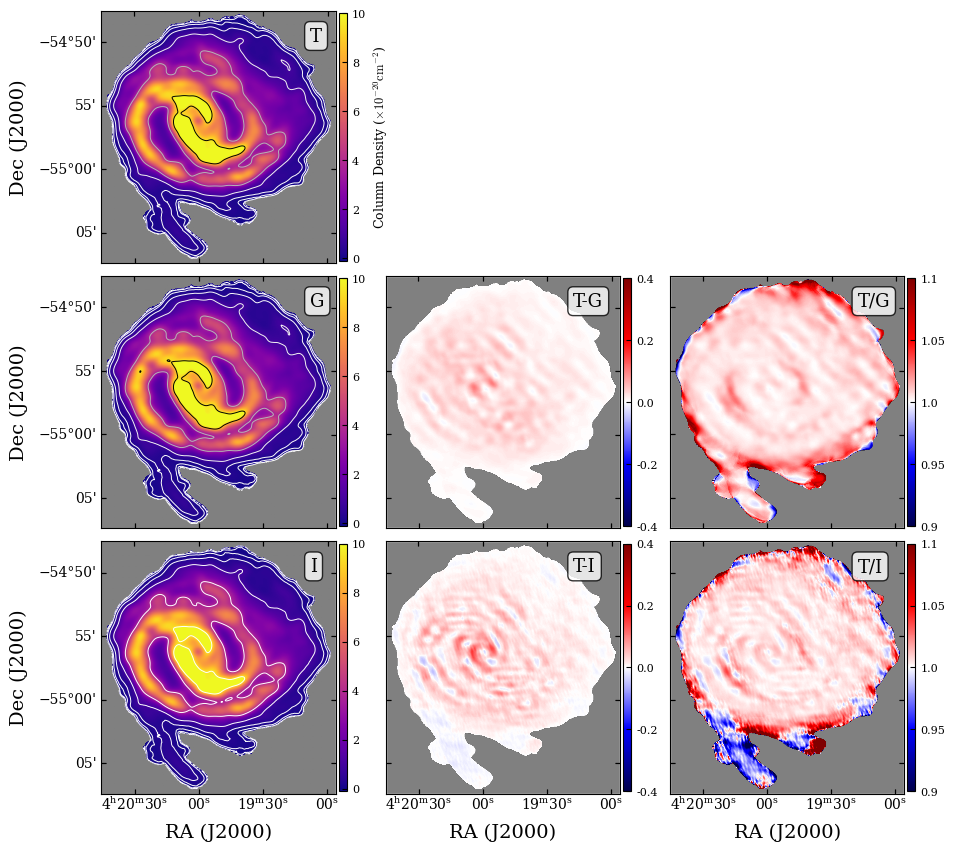} %plot_aligned_mom_maps_r20_corr.png}
    \caption{Moment 0 images at 40\arcsec{} resolution and Natural weighting (Robust 2.0), of NGC\,1566 from SOFiA, using (left column) traditional processing (T), Grid-stacking (G) and Image-stacking (I). The middle column is the difference between traditional and the other imaging methods. The right column is the ratio.
    The moment-0 maps all are visually similar, {and the reconstruction of high spatial frequencies, as indicated by the differences and ratios, are very close for both the Stacking methods.}
    \label{fig:image_compare_2.0}}
\end{figure}

\begin{figure}
    \centering
    \includegraphics[width=0.48\linewidth]{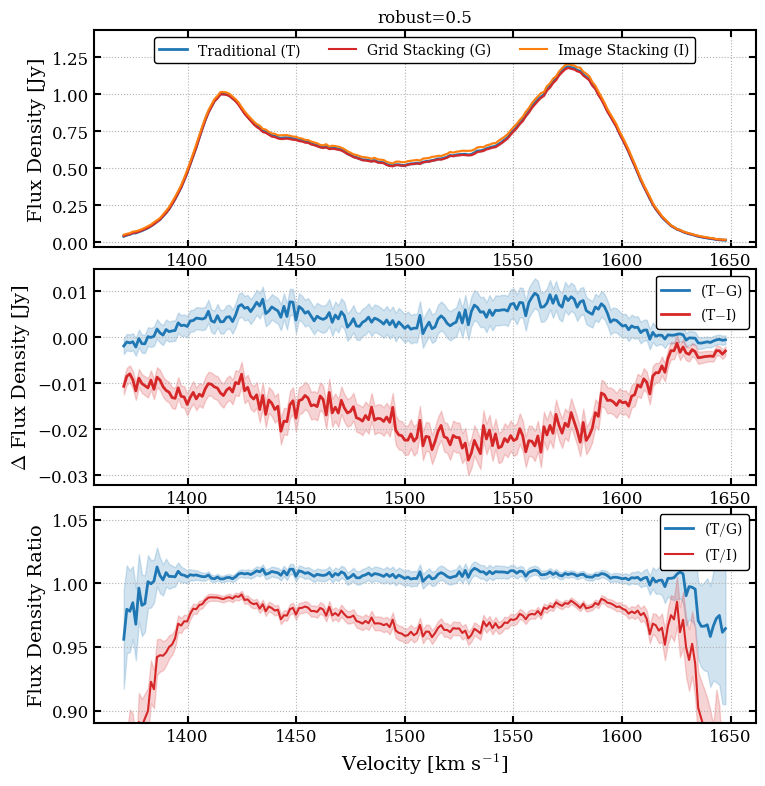}
    %plot_compare_spec_deep_imaging_r05_corr.png}
    \includegraphics[width=0.495\linewidth]{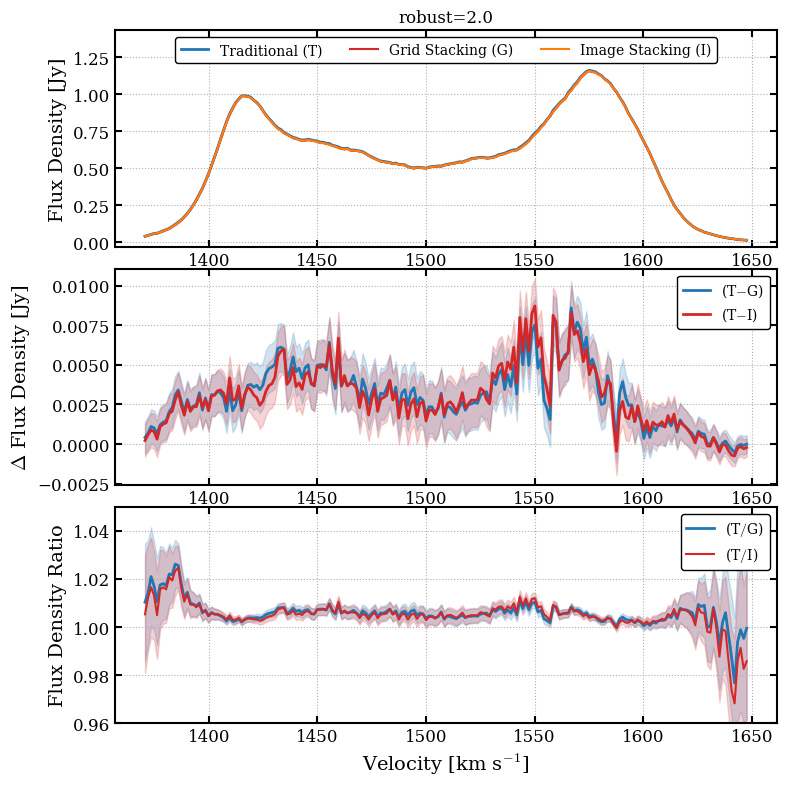} %plot_compare_spec_deep_imaging_r20_corr.png}
    \includegraphics[width=0.48\linewidth]{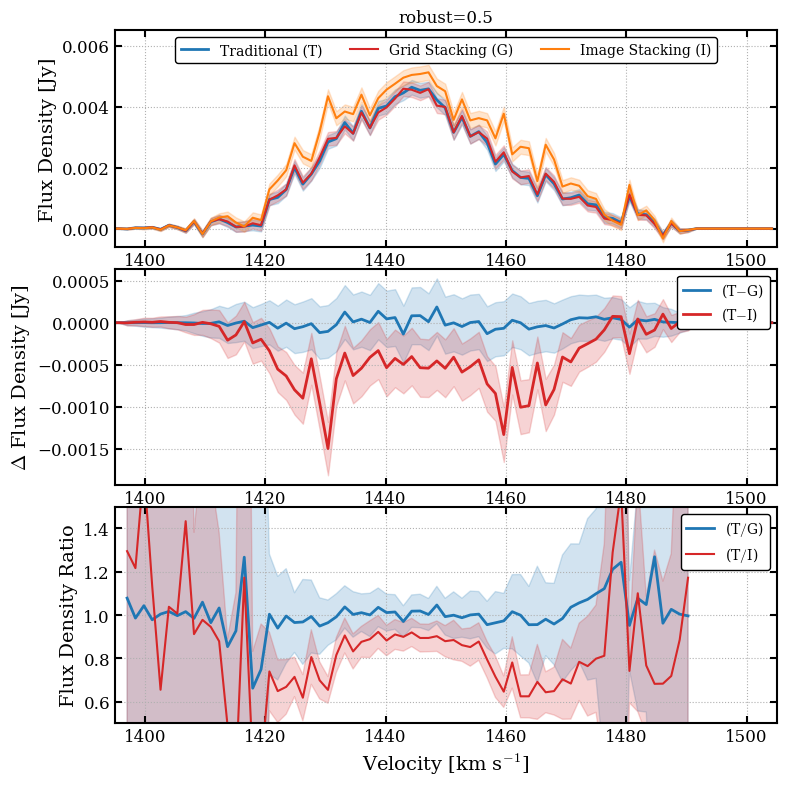}
    %plot_compare_spec_deep_imaging_faint_source_r05_corr.png}
    \includegraphics[width=0.48\linewidth]{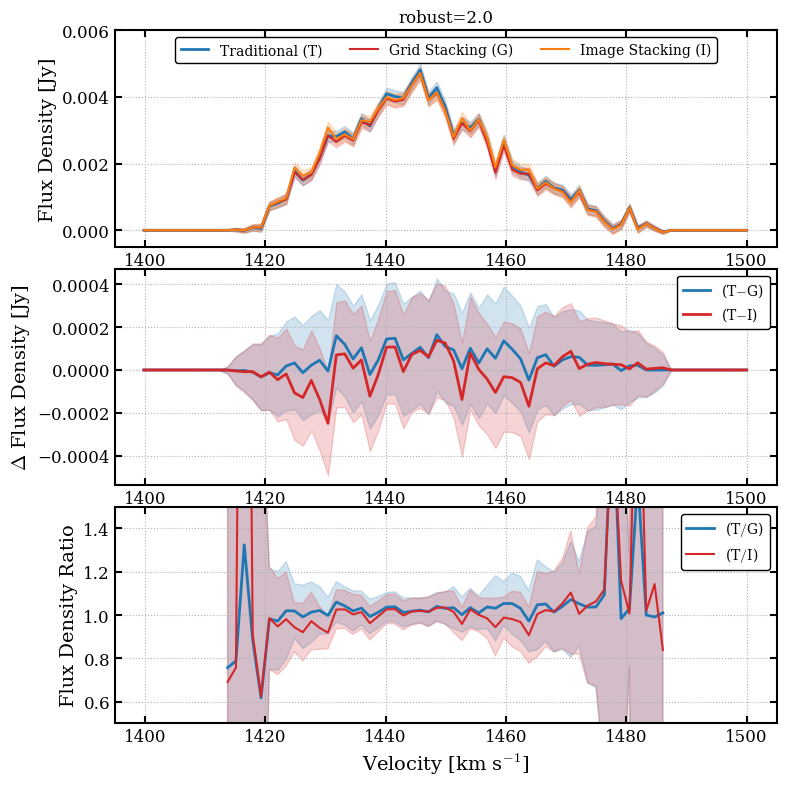}
    %plot_compare_spec_deep_imaging_faint_source_r20_corr.png}
    \caption{The spectra of NGC\,1566 (top) and for J0422--5455 (bottom) at Robustness 0.5 (left) and 2.0  (right) from SOFiA with shaped beams of 20\arcsec{} and 40\arcsec{} respectively, 
    using traditional processing (T) as a reference and compared with the spectra from Grid-Stacking (G) of the cubes and from Image-Stacking (I) of the cubes. 
    {We find that, for the higher resolution Robust 0.5 imaging, the stacking methods do not perfectly reproduce the traditional processing, with the grid-stacking finding about 0.7\% less flux and the image-stacking finding about 2.7\% more flux on this strong and well-modelled source.
    On the other hand, for the natural weighted image, the reconstruction with both stacking methods agree with each other and finding about 0.6\% less flux than the traditional processing.
    }
    %\RD{Robust 2 grid stacking needs remaking, So these will be replaced.}
    %\JR{\bf Both grid-stacking and image-stacking to be redone }
    }
    \label{fig:spectra_compare}

\end{figure}

The comparison of the final data product, that being the spectra and moment zero maps (Figures \ref{fig:image_compare_0.5} to \ref{fig:spectra_compare}), and the derived values are shown in Table \ref{tab:ngc1566_params}.
The results for the images at Robust 0.5 are shown in Figure \ref{fig:image_compare_0.5}, which has the smaller beam size of 20\arcsec{}, %show agreement between the different approaches.
and Figure \ref{fig:image_compare_2.0}, 
%\RD{in the appendix ?} 
shows the same but for natural weighting. 
Natural weighting has 40\arcsec{} beam size and show excellent agreement between all the methods with and without beam-shaping, %for NGC\,1566, 
supporting our suspicion that differences arise when the restoring beam is poorly estimated.
%the results from the strongest source were compatible across all methods, and it was in the weaker sources that a difference in the quality of the results from image or grid-stacking could be identified. 
%
The results for the spectra with natural weighting and Robust 0.5 are shown in Figure \ref{fig:spectra_compare}.
{The mean difference for the central half of the spectra (1450--1580\kms) between grid-stacking and traditional processing is -0.7\% at Robust 0.5 and -0.6\% at Robust 2.0. 
In comparison, the mean difference between image-stacking and traditional processing is +2.7\% at Robust 0.5 and -0.6\% at Robust 2.0.
Our results are compatible to those from \citet{rhee_25} where  image-stacking fell short of grid-stacking, when compared with traditional processing with a robustness of 0.5.
As a note of caution, we point out that in \citet{rhee_25} image-stacking for the strong source showed good performance, whilst for the weaker sources the recovered flux fell short, and we have only one weak sources in this dataset.
}
%     \centering
%     \includegraphics[width=0.6\linewidth]{plot_aligned_mom_maps_20km.png}
% \end{figure}
% plot_compare_spec_deep_imaging_r05_t20_new_mask.png
% plot_compare_spec_deep_imaging_faint_source_r05_t20_new_mask.png
% plot_aligned_mom_maps_r05_t20_new_mask.png
% plot_aligned_mom_maps_faint_r05_t20_new_mask.png
% plot_aligned_mom_maps_r20_t40_new_mask.png
% plot_compare_spec_deep_imaging_r20_t40_mask.png
% plot_compare_spec_deep_imaging_faint_source_r20_t40_new_mask.png
% plot_aligned_mom_maps_faint_r20_t40_new_mask.png

\begin{table}
\caption{Comparison of the physical parameters of NGC\,1566 (at Robust 0.5), from the MHONGOOSE analysis (labelled WSc) and our results derived by SoFiA on the the images generated by the different methods: Traditional processing (T), Grid-Stacking (G), Image-Stacking (I) and Traditional processing of DYSCO data to 4-bits compressed (D4), MGARD data compressed to errors of 0.3\,Jy (M4) and  SZ data compressed to errors of 0.3\,Jy (S4).  
All science outcomes are within errors. The distance to NGC\,1566 is taken as 17.7Mpc as in \citet{anand_21}. 
\New{%referee: One minor revision: - please add the peak of the residual in table 2, alongside the rms.
The Peak Residual is taken from the residual cubes, as SoFiA does not report this value.}
}
\label{tab:ngc1566_params}
    \centering
\begin{tabular}{l|r|rrr|rrr}
\toprule
Parameter & WSc & T & G & I & D4 & M4 & S4\\
\midrule
RMS  [mJy~beam$^{-1}$] & 0.171 & 0.166 & 0.166 & 0.171 & 0.169 & 0.168 & 0.168  \\
% These are taken from the best sources I can find. For the majority of ours I used resid*4-max.casa
\New{Peak Residual  [mJy~beam$^{-1}$]} & 0.83 & 0.76 & 0.79 & 1.36 & 0.83 & 0.80 & 0.81  \\
S$_{\rm int}$ [Jy~Hz] & {762674.14} & 781123.80 & 770511.84 & 759978.61 & 781226.44 & 781208.34 & 781196.83 \\
$W_{20}$ [{\kms}] & {220.59} & 222.23 & 222.72 & 222.61 & 222.28 & 222.25 & 222.22 \\
$W_{50}$ [{\kms}] & {210.98} & 212.22 & 212.23 & 212.27 & 212.21 & 212.22 & 212.20 \\
log ({\MHI}/{\msun}) & {10.06} & {10.09} & {10.08} & {10.07} & {10.09} & {10.09} & {10.09} \\
\bottomrule
\end{tabular}
\end{table}

\subsection{Comparison of traditional imaging with and without lossy compression}
%%%%% Comparison of traditonal processing without (dt) and with (dc) compression %%%%%%%%%

Due to the processing time for the full dataset, imaging for only two of tested compression error bounds was attempted, which were selected based on the impact on the compression on the visibility data, see Figure \ref{fig:eb_meerkat}. 
We used our docker demonstrator \texttt{docker-casacore} to compress the data. 

For ADIOS we tested both the SZ and MGARD compression algorithms, with an absolute
error bound of 0.3 and 0.03\,Jy (about 20\% and 2\% of the \Old{system equivalent flux density (SEFD) (395/$\sqrt{6,500\times8.1}$\,Jy) for these integrations and bandwidths}.  For MGARD this gives a compression to $\sim$12\% and $\sim$22\% of the original size, respectively, and for SZ to $\sim$14\% and $\sim$24\% of the original size. 
The 20\% error bound is beyond what we have previously recommended \citep{dodson_25}, particularly as we are forming a  final product that is three times deeper than the daily image, but was selected to ensure that there were detectable impacts.
We analysed the data in the same way as the traditional processing, but accessing the data after compression with ADIOS.
Figure \ref{fig:eb_meerkat} plots the impact of a range of error bounds on the compression ratio and the  maximum and rms errors introduced by compression of the visibility data, as a fraction of the standard deviation (1.8\,Jy in this case).

\begin{figure}
    \centering
    \includegraphics[width=0.8\linewidth]{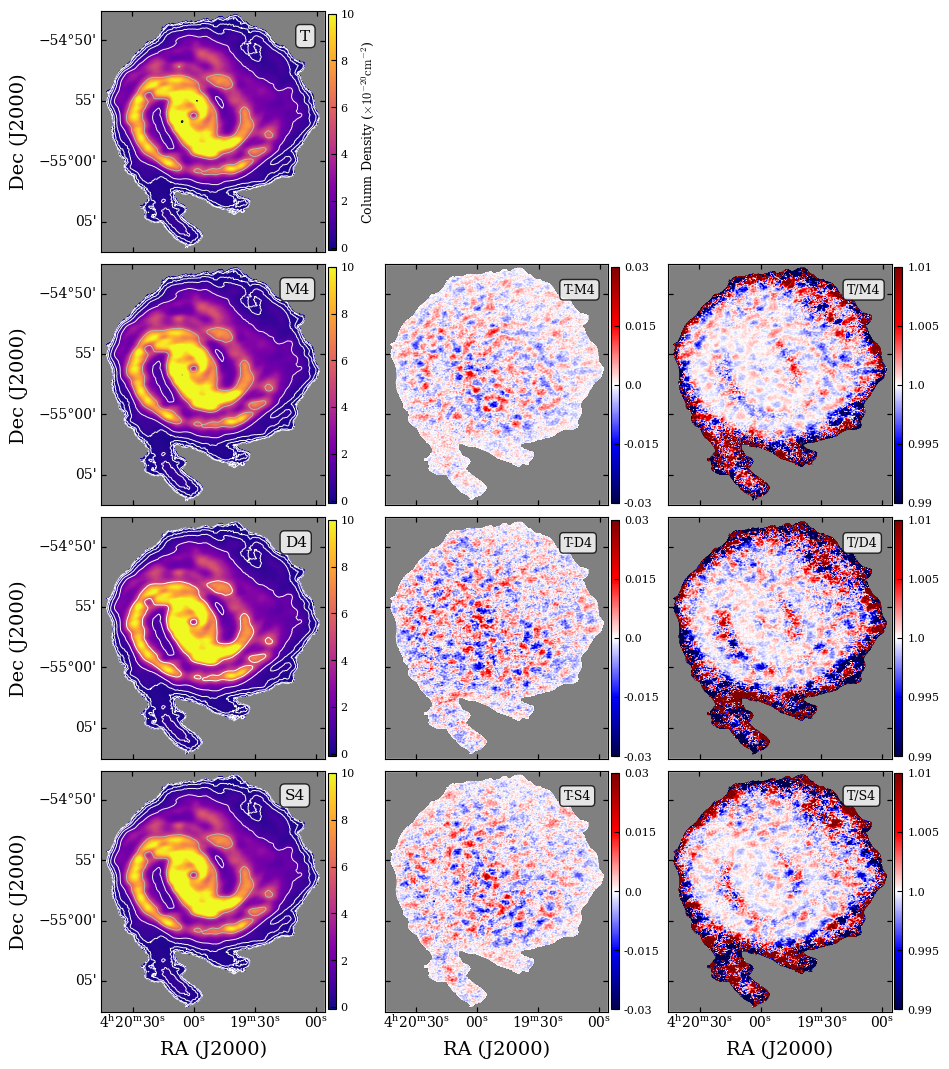}
    \includegraphics[width=0.48\linewidth,trim={0.4cm 0.4cm 0 0},clip]{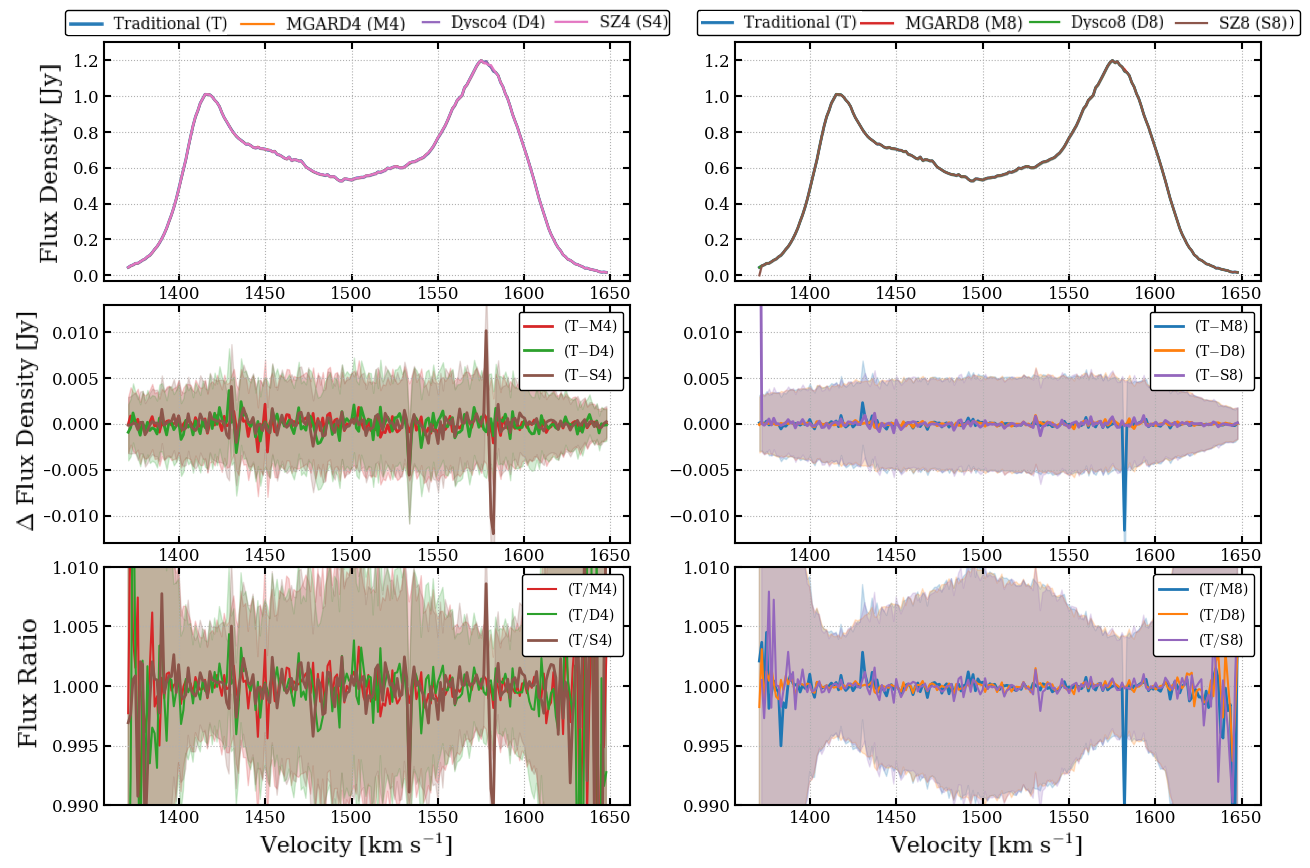}
    \includegraphics[width=0.48\linewidth,trim={0.4cm 0.4cm 0 0},clip]{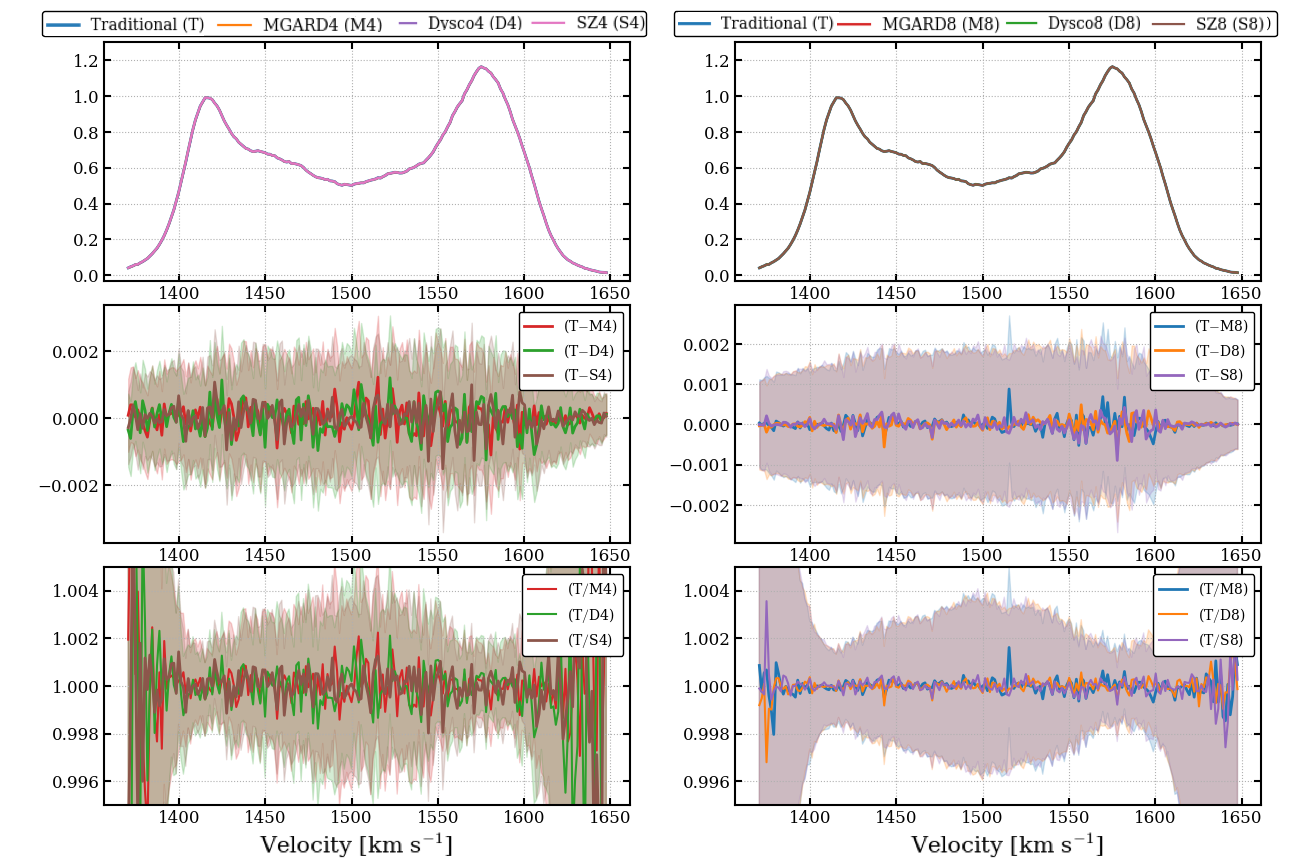}
    \caption{Comparison of the images and spectra from SOFiA using traditional processing, without and with compression. 
    The images (top) show the moment-0 map at (for Robust 0.5) using non-compressed data, or data compressed with MGARD, DYSCO or SZ, at 4-bits or $\sim$22\% levels (labelled accordingly e.g. M4, D4 and S4), with differences and ratios in middle and right column respectively. 
    The imaging results for the less aggressive compression was better and is not shown.
    The reconstruction of high spatial frequencies, as indicated by the differences and ratios, are close to featureless.
    The comparison of the spectra (bottom) (left: Robust 0.5, right: Robust 2.0) of the images made traditionally and with 4-fold and 8-fold compressed data, respectively, underline the quality of the results with this approach; 
    the outputs are indistinguishable to within errors (shaded band). This is unsurprising as the compression methods guarantee that the mean values are preserved, and
    %an image can be thought of as the mean value over all baselines for many directions. Furthermore, 
    the clipping used in the formation of the data products ensures only pixels with strong emission are included in the analysis.
    \label{fig:spectra_comp_compare}}
\end{figure}

For DYSCO we
chose an encoding level of 8-bits and 4-bits to give an four-fold and eight-fold compression, and to approximately match the ADIOS compression.
We analysed the data in the same way as the traditional processing, but accessing the data column compressed with DYSCO.
The top of Figure \ref{fig:spectra_comp_compare} plots the comparisons for three traditionally processed trials, with both Robustness 0.5 and 2.0, using non-compressed data, ADIOS-compressed data and DYSCO-compressed data where the data volume is reduced by a factor of exactly eight for DYSCO (4-bits representation; D4) or approximately eight for SZ and MGARD (S4 and M4, respectively). 
The moment-0 maps show errors on the scale of $\sim$1\% in difference and $\le$1\% in ratio, even with an order of magnitude of compression.
%{\bf what is the error bound here - why are the scales are 4-bit and 8-bit so different}
The bottom of Figure \ref{fig:spectra_comp_compare} plots the comparisons for the seven traditionally processed trials (no-compression, compressed in size to $\sim$12\%: M4, D4 and S4, and $\sim$20\%: M8, D8 and S8), with both Robustness 0.5 and 2.0. 
For the ADIOS compressors and with the error bound at 20\% of the SEFD, the difference in the inner half of the spectra is less than $\sim$0.01\%, with no significant structure over the difference or ratio of moment zero images. With the error bound at 2\% of the SEFD, the difference is less than $\sim$0.001\%,  with no significant structure.
For the DYSCO compression and an eight-fold reduction in size the difference in the spectra is less than $\sim$0.01\%, with no significant structure over the difference or ratio of moment zero images. With a four-fold reduction in size the difference in the spectra is about $\sim$0.001\%, without significant structure.

\subsection{Comparison of images reconstructed with different weights}

\begin{figure}
    \centering
    \includegraphics[width=0.7\linewidth,trim={0 10.5cm 0 0},clip]{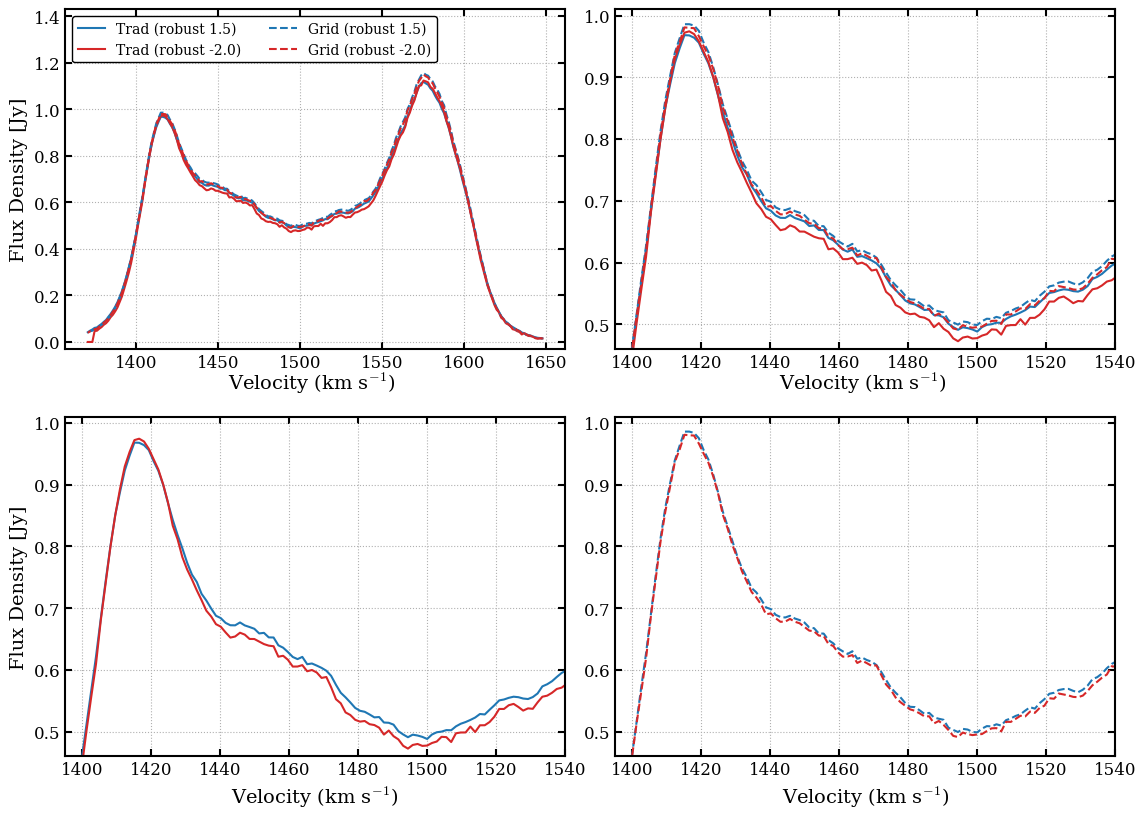}
    \caption{The profile of the galaxy with Robust weights -2.0 (red) and 1.5 (blue), made traditionally (solid lines) and with re-weighted Grid-Stacking (dash lines). 
    The grid imaging was via reprocessing the natural weighted (Robust 2.0) grids. 
    Left shows the full spectra and Right shows a zoomed portion.
    The high resolution uniform-weighted spectra and the lower resolution Robust 1.5 spectra, both reconstructed from the natural weight grids, closely match that of the naturally weighted spectrum. Ob the other hand, the spectra from the traditionally imaged uniform-weighted cubes show significant missing flux, but the traditional Robust 1.5 spectra shows good flux recovery compared to the natural weighting. 
    We postulate that this is because the initial daily imaging at natural weighting has lead to a deeper subtraction of the extended flux, which is missed in the traditional imaging.
    %Bottom: the moment zero images at Robust -2.0 and 1.5 for Traditional processing and from reconstruction from the natural grid.
    \label{fig:reweight}}
\end{figure}

{If one has formed and retained the cube grids one can reconstruct this data with a different weighting scheme, with a final minor CLEAN to adjust any differences in the derived models from those one would have obtained via traditional processing. 
To demonstrate this, we imaged the data traditionally with a Briggs Robustness of 1.5, and -2.0 (uniform). 
These were compared to the matching image made from the Robustness 2.0 weighted (natural) grids, with a minor CLEAN of the summed residuals at the desired weighting (1.5 or -2.0) and the restoration of the pre-existing models with the correct beam-size into that restored residual image cube.
} 
These steps ensures that any imperfections in the reconstruction from the multiple shallower and differently weighted models are corrected for in the final step.
{For example, the broadscale structure not picked up by the traditional processing at Robust 1.5, that was detected and modelled at Robust 2.0, 
clearly appears in the top row of Figure \ref{fig:reweight-image}. 
The bottom row plots the comparison images for grid-reconstruction and traditional processing at uniform-weighting, where the difference in robustness between the grid weighting and the reconstruction is as large as it could be (2.0 to -2.0).} 
The traditional processing time for re-imaging with a different weight was $\sim$30hours whereas to reconstruct the cube with a different weight from the daily grids was $\sim$5min.
Figure \ref{fig:reweight} plots the spectra, where similar performance can be observed. The grid-reconstruction has a similar profile for all weighting schemes, whereas the traditional processing has a similar profile, but not identical.

{Our explanation is that the images reconstructed from the natural-weighted grids are slightly deeper, having picked up more extended flux in the modelling at natural weighting, than the images made at the higher resolution and lower robustness. This, one could argue, means that the reconstructions are closer to the truth on the sky, but our metric was that they reproduce the traditional processing, on which basis one would conclude that they over predicting the spectra by the order of $\sim$10\%.}
%
%It is likely that scales for the  natural and uniform model fitting (0,6,15,30) have an overlap. 
%}

\section{Conclusions}
%%%%%%%%%%%%%%%%%%%%%%%%%%%%%%%%%%%%%%%%%%%%%%%%%%
% We then can:
%   * Make a daily image and sum those for a grid-stacking demonstration
%   * explore how we apply different imaging weights and tapers on a master grid 
%   * Sum the images for a image-stacking demonstration
%   * Image all the MS files for a traditional demonstration
%   * Compress the MS files and then image all those for an alternative traditional demonstration
%%%%%%%%%%%%%%%%%%%%%%%%%%%%%%%%%%%%%%%%%%%%%%%%%%

% These are the conclusions I see:

% MS files are ~24GB (or 1/2 of that?)
% grids are: 1.8*6
% ~0.4GB after compression (80%)
% images are 3.2GB
% Times from log files, compression time is *10 for 10 MS
\begin{table}
    \centering
    \begin{tabular}{|c|rcccr|}
    \hline
      Strategy   & Error Bound & Fractional Compression & Compute & Image Impact & Science Impact  \\
         &  &  (\%)& 201$\times$hrs & ($\sigma$) RMS/Max & (\%) \\
      \hline
      \hline
      &\multicolumn{5}{c|}{Daily Clean Only: Long-term solution with staged compute}\\
      \cline{2-6}
      Image-stacking  & Lossless& 13$^\dag$ &10$\times$9.9 & 0.28/16 & 2.7 \\
      %\hline
      \hline % \cline{2-6}
      &\multicolumn{5}{c|}{Hogbam Clean Only: Long-term solution with staged compute}\\
      \cline{2-6}
      Grid-stacking  & Lossless& 10$^\ddag$&10$\times$9.9 & 0.10/4.7 & 0.7 \\
      %\hline
      \hline % \cline{2-6}
      Lossy:&\multicolumn{5}{c|}{Major Cycle Clean: Not compatible with current planned ASKAP or SKA compute}\\
      \cline{2-6}
      DYSCO  & 4-bit  & 12.5$^*$& 34+0.4 & 0.15/1.6& 0.01\\
      DYSCO  & 8-bit  & 25$^*$& 34+0.4 &0.01/0.7& 0.001\\
      MGARD  & 20\%\,SEFD& 12$^*$ &34+3.6 & 0.10/1.6& 0.01\\
      MGARD  & 2\%\,SEFD& 22$^*$ &34+3.6 & 0.01/0.8& 0.001\\
      SZ  & 20\%\,SEFD& 14$^*$ &34+1.0 &0.12/1.2& 0.01\\
      SZ  & 2\%\,SEFD& 24$^*$ &34+1.0 &0.06/7.1 & 0.001\\
    \hline
    \end{tabular}
    \caption{\Old{Comparison of strategies (Column 1: image-stacking, grid-stacking and lossy compression of raw data) for processing deep spectral line data in the SKA-Era. 
    Column 2 reports the error bound, either loss-less, a bit-limit (DYSCO) or the visibility error bound compared as a fraction of the SEFD.
    Column 3 reports the reduction in the data-size; image-stacking and grid-stacking are compared the size of the input MS. 
    Column 4 are estimates for the compute required. For the stacking process each individual epoch is processed as it arrives. For the lossy strategies the given time is the traditional imaging time plus the time to compress and uncompress the data columns for every epoch, using the slower CPU implementation. 
    Column 5 are the maximum absolute (peak) and RMS differences to the traditional processing, as a fraction of image RMS (0.2mJy/beam) for the Robust 0.5 image cube.
    Column 6 is fractional error in the science output on the galaxy parameters, from the Robust 0.5 cube.\\    
    $^\dag$ Restored Image to MS size; $^\ddag$ compressed grid products to MS size; $^*$ compressed data column compared to uncompressed column in MS.}}
    \label{tab:conclusions}
\end{table}

In this study, we evaluated next-generation strategies for deep spectral line imaging using data on one galaxy from the MeerKAT MHONGOOSE survey, focusing on approaches to mitigate the storage and processing constraints anticipated for future large-scale radio instruments. 
{We firstly confirmed that our results were internally consistent and secondly consistent with previous work by the MHONGOOSE team.
By comparing traditional imaging in \askapsoft{} against image-stacking, grid-stacking, and lossy compression techniques, we have identified several viable pathways for data reduction in the SKA-era where storage costs will significantly limit to processing options.
By comparing the results with those from WSClean we confirm that the ASKAPSoft results are compatible with this alternative analysis path, once the well-known consideration of the definition of Stokes I was taken into account.}

Our findings continue to highlight grid-stacking as a promising method. 
{It closely reproduces the traditional processing results (to $\sim$0.7\% at Robust 0.5), whilst suppressing imaging artifacts associated with uncleaned emission.
Image-stacking does reproduce the traditional results reasonably well (to $\sim$2.7\% at Robust 0.5), but will suffer from artifacts from uncleaned emission now visible in the deeper cube. 
At Robust 2.0, with a larger beam, imaging the errors for both image-stacking and grid-stacking are close to equal at $\sim$0.6\%.}
Whilst the computational requirements for both stacking approaches remain comparable to traditional approaches, stacking offers the distinct advantage of distributing processing requirements over the duration of an experiment through daily imaging. 
We do not expect to be able to process SKA-scale datasets in the traditional processing; it would already be impossible for the full 800h of DINGO.
In the DINGO pipeline the grid data products come `free' with the normal data reduction.
Furthermore, this technique enables the reconstruction of a full range of tapers from a single data reduction -- provided additional major cycles are not required -- offering significant flexibility for multi-scale science. 

However, further refinement regarding the determination of the correct beam size during restoration is necessary to ensure scientific consistency.
We used the simple solution of using a shaped beam to ensure that every epoch reconstructed the image with a constant beam shape over the experiment, as we do for DINGO, to remove these issues. 
In this dataset we do not have a number of weaker targets to provide a meaningful comparison of grid-stacking against image-stacking at low flux levels, as we did for DINGO. 
This should be attempted in future work, where one can include all the sources in the field and frequency range, but this was not goal of these investigations to demonstrate that grid-stacking was a viable solution for MeerKAT data.

Additionally, we demonstrated that traditional imaging using heavily compressed data yields results that are {near-perfect (the fractional difference between the spectra is $<\sim$0.01\% for in all cases}), suggesting that lossy compression is a robust alternative solution for managing data volumes. 
This introduces fewer errors, and is closely aligned with the traditional processing methods and we expect this to be a popular option. However, we would raise the concern that the resources required to fully image full multi-epoch datasets at the end of observations could be a significant challenge, and it maybe that the only practical approach is to spread it out in parts, such as for the grid-stacking.
Certainly with setonix we would not be able to traditionally image in one pass the 800h of DINGO data expected by the end of this year. 
{Furthermore, we note that our analysis of ten datasets reached a final depth only three times that of a single epoch, and we would recommend more extensive testing is required before committing to this solution.
Future work leveraging larger datasets, such as the 400-hour DINGO survey, will be essential to fully validate these compression and stacking strategies for the next generation of radio interferometry.}

\Old{These performance studies are drawn together in Table \ref{tab:conclusions}, where the two methods and 6 levels of lossy compression are compared. 
All of these compression approaches introduce small errors in the science and image cube data products. 
For the grid stacking we introduce about an increase of 10\% in the noise floor in image cube.
The impact on the science outcome, being the summation over all the pixels in the image cube, is less than 1\%. 
The compute is spread over the duration of the project (here being 10 epochs, but for DINGO this will be about 400 individual epochs). 
Based on the current compute infrastructure we estimate that we will not be able to process the required 800 hours of data from each DINGO tile and band in a single processing run. 
However, for smaller projects such as MHONGOOSE, traditional processing is still possible. The total time for imaging on setonix was about 10 hours per epoch as opposed to 34 hours for traditional processing.
Note that no effort was made to optimise these, so they are upper bounds.
For the lossy compression followed by traditional processing, there is about an increase of 10\% in the noise floor in the worst cases, which also reduces the the data volume size by a factor of 10, or up to a 1\% increase in the noise floor and an decrease in the data volume by about a factor of 4 for DYSCO-8 bit or MGARD-2\%. 
The maximum error (around the strongest emission) is the order of a few times the noise floor.
The impact on the science outcome is minimal, as this approach allows for a major cycle clean.

In summary, where long-term storage is an issue but compute will not be (such as MHONGOOSE), lossy compression the data column is an attractive solution as it allows for traditional processing with major cycles at the end of the project.
Where sufficient compute for the multiple epochs will not be available, spreading the effort over the lifetime of the project and storing intermediate products is the only alternative to the image-stacking solution. 
Image-stacking would be acceptable for a limited number of epochs; a 3$\sigma$ uncleaned residual in the cube would become a 10$\sigma$ uncleaned feature in a stack of just 10 epochs. Such features would affect the science outcomes, with the degree of impact depending on the PSF sidelobe levels.
Thus, for DINGO with 100 epochs per pointing, grid-stacking remains our best solution} \New{and for MHONGOOSE with 10 epochs, lossy-compressing the data would be the best solution.
}

\section*{Acknowledgements}
This research used the resources of the Pawsey Supercomputing Research Centre, {under allocations JA3 and PAWSEY0411}. Establishment of 
the Pawsey Supercomputing Research Centre was an initiative of the Australian Government, with support from the Government of Western Australia and the Science and Industry Endowment Fund. 
Part of this work has received funding from the European Research Council (ERC) under the European Union’s Horizon 2020 research and innovation programme (grant agreement No 882793 ‘MeerGas’).
Part of this research was supported by the Pawsey PACER project. 
This research made use of 
Matplotlib \citep{Hunter:2007}, and Numpy \citep{Harris:2020}.

\section*{Data Availability}

% The data underlying this article will be shared upon reasonable request to the MHONGOOSE team, via EB. \New{(?)}\EB{
% Data cubes are available via mhongoose.astron.nl. But not the measurement sets. The full ms measure ~10 TB and there is no convenient way to distribute these - plus they can be obtained from the MeerKAT archive)}
MHONGOOSE data cubes are available via mhongoose.astron.nl, and the raw measurement sets can be obtained from the public MeerKAT archive.
The \askapsoft{} software suite and the docker for compression are publicly available. 

\paragraph{Author Contributions}
% Please provide an author contributions statement using the CRediT taxonomy roles as a guide {\verb+\url{https://www.casrai.org/credit.html}+}. Conceptualization: A.A; A.B. Methodology: A.A; A.B. Data curation: A.C. Data visualisation: A.C. Writing original draft: A.A; A.B. All authors approved the final submitted draft.
Writing original draft and figures: R.D., M.R., J.R.; PI of MHONGOOSE: W.B.; PI of DINGO: M.M.; Developers of/contributors to grid-stacking code: A.W., K.R., D.M., P.E.; All authors approved the final submitted draft.



%\bibliographystyle{mnras}
%\bibliography{references_deep_imaging} % if your bibtex file is called example.bib

%\begin{thebibliography}{99}
%\end{thebibliography}

%%%%%%%%%%%%%%%%%%%%%%%%%%%%%%%%%%%%%%%%%%%%%%%%%%

%%%%%%%%%%%%%%%%% APPENDICES %%%%%%%%%%%%%%%%%%%%%

\clearpage
\appendix
%\pagebreak
\section{Additional details}
\begin{figure}%[!h]
    \centering
    \includegraphics[width=0.5\linewidth,trim={0 0.cm 0cm 0.cm},clip]{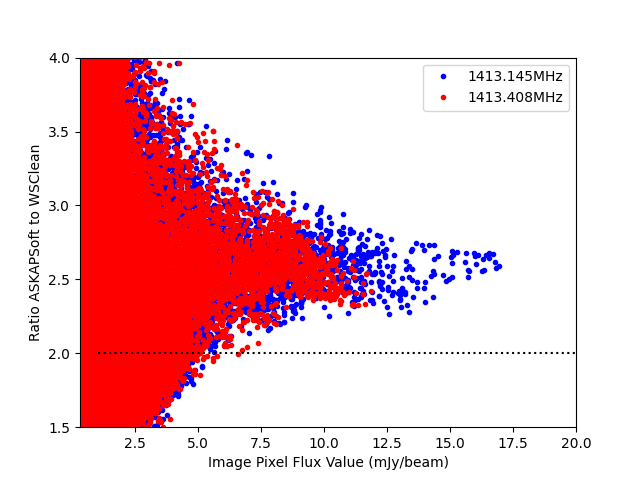}
    \caption{Ratio of \askapsoft{} to WSClean pixel flux values \Old{in an example data cube, before correction for the definitional difference of a factor of two, for two channels in the image cube. ASKAPSoft has a slightly higher than expected flux ratio, and a marginally increased noise floor, which appears to be due to the different implementation of the multiscale clean parameters between the two imagers.}}
    \label{fig:ratio_flux}
\end{figure}
\New{Figure \ref{fig:ratio_flux} provides the pixel to pixel comparison for the Stokes-I image cubes from WSClean and from ASKAPSoft, demonstrating the difference in the fundamental definition in the formation of Stokes-I, that leads to a scaling factor of 2. 
The additional factors we associate with the different multi-scale algorithms, as the difference reduces (but not to the point of being identical) when the same nominal multiscale factors are used.
%Following this investigation we decided 
We choose
to correct the ASKAPSoft pixel values by a factor of two, and to compare our new images to traditional images made in ASKAPSoft rather than compare these directly to the published WSClean images.
}

\begin{figure}%[!h]
    \centering
    \includegraphics[width=0.5\linewidth]{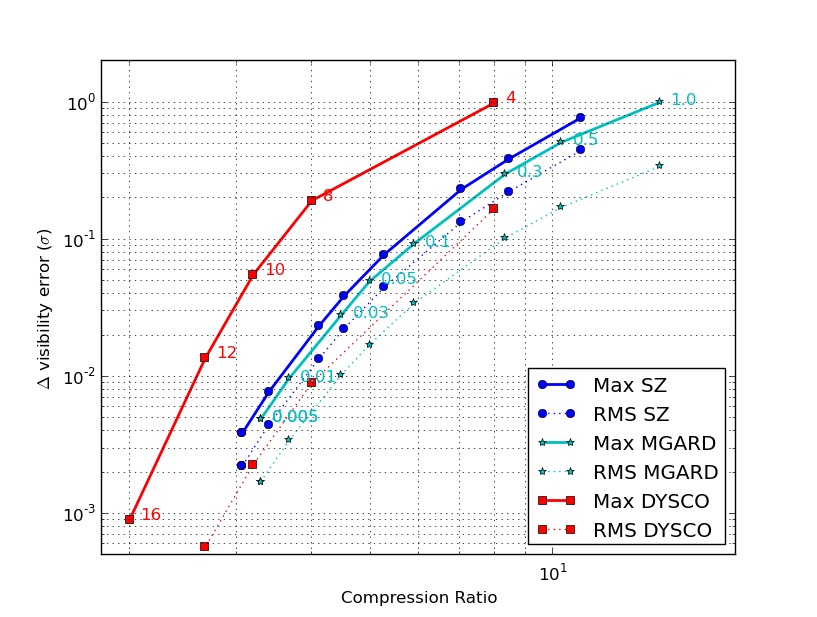}
    %ADIOS_Compression_Errors.MeerKAT.png}
    \caption{Degree of compression and the impact in RMS and maximum error in the difference of the visibility data, as a function of error bound for the SZ (blue circles) and MGARD (cyan stars) compressors, and as a function of bit level for the DYSCO compressor (red squares). Both maximum error (solid lines) and RMS error (dotted lines) as a fraction of the visibility standard deviation are plotted.
    \Old{This can be used to estimate sensible levels of compression to apply.}
    For the ADIOS compressors the individual points are labelled with the requested error bound; for DYSCO the points are labelled with the bit limit.
    }
    \label{fig:eb_meerkat}
\end{figure}
\New{Figure \ref{fig:eb_meerkat} shows the impact on the visibility thermal noise (derived from the RMS of the difference in the visibility values with and without compression) as a function of the compression error bound. 
The tighter the error bound (or equivalently the number of bits for DYSCO) the smaller the impact on the thermal noise. A factor of ten in compression, for example, leads to about a 20\% increase in the noise for MGARD, and about a 30\% increase for both SZ and DYSCO. 
The maximum values are greater than the RMS difference, and significantly greater for DYSCO. These lead to our choice of MGARD as our default compressor, even though it requires the greatest compute. The impact on the image cubes and the science outputs, rather than the visibilities, is significantly less as these are the average of many visibilities. }

\begin{figure}%[!h]
    \centering
    \includegraphics[width=0.8\linewidth]{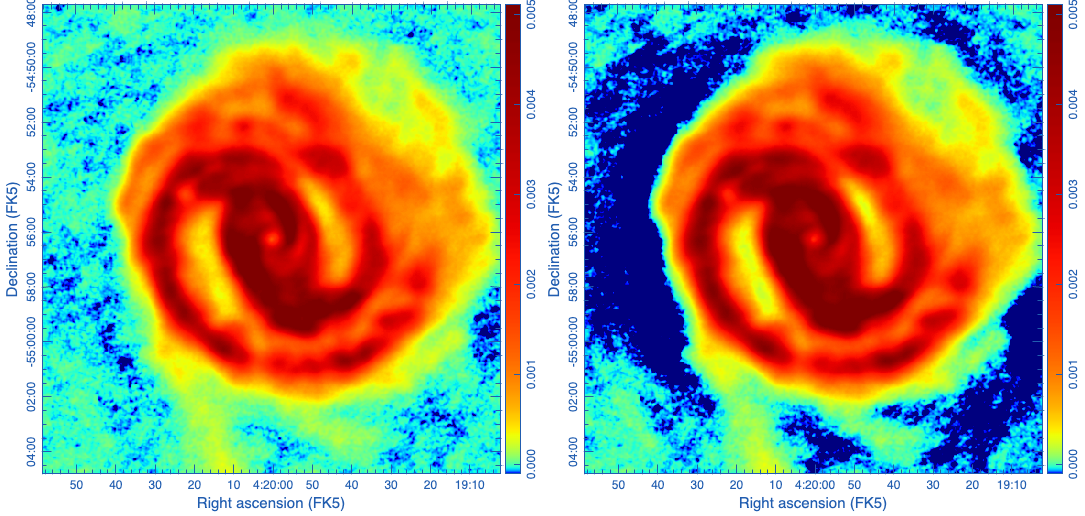}
    %Screenshot 2026-05-06 at 16.37.16.png}
    \includegraphics[width=0.8\linewidth]{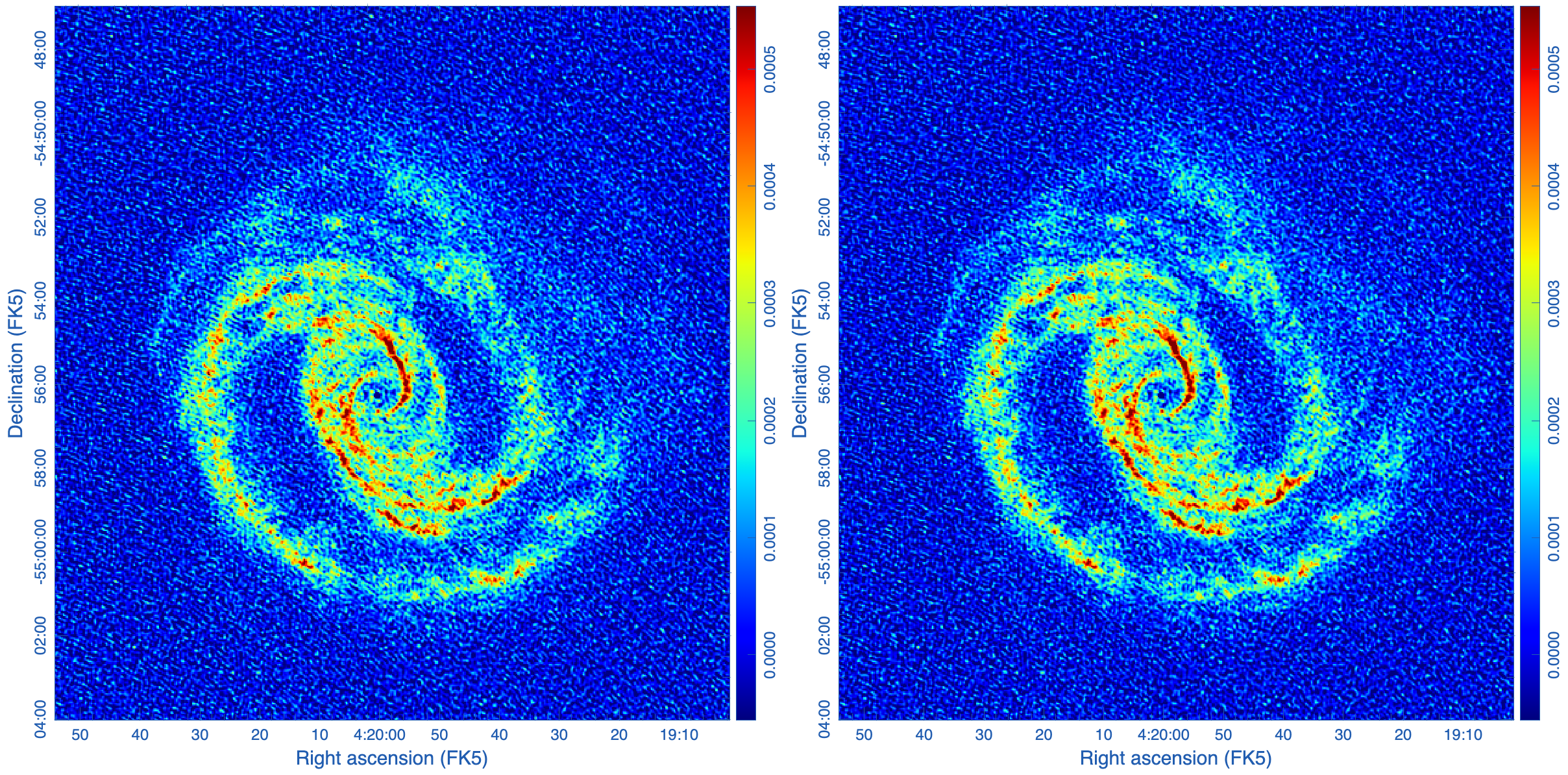}
    %Screenshot 2026-05-06 at 16.38.10.png}
    \caption{Comparison of the reconstructed images (left) compared to traditional imaging (right), for (top) 
    Robust 1.5 and (bottom) Robust -2.0 (uniform), when averaged over all channels. 
    In the stretched log colour scale one can see the negative bowl left by the traditional processing at the lower resolution, where there is less capacity for extracting broad scale features. 
    \label{fig:reweight-image}}
\end{figure}
\New{Figure \ref{fig:reweight-image} shows some examples of the moment-zero maps generated from the naturally weighted grids. 
Their similarity suggests that there are no barriers to making image cubes with multiple angular sensitivity scales from uv-grids, without significant computing costs. 
Further work is required to confirm this promising path.}

%\section{Processing Parameters}

% If you want to present additional material which would interrupt the flow of the main paper,
% it can be placed in an Appendix which appears after the list of references.

%%%%%%%%%%%%%%%%%%%%%%%%%%%%%%%%%%%%%%%%%%%%%%%%%%

% Don't change these lines
\bsp	% typesetting comment
\label{lastpage}
\end{document}